\documentclass[amsmath,aps,showpacs,a4paper,10pt]{revtex4}

 \usepackage{epsf}
 \usepackage{graphicx}    

\begin{document}

 \newcommand{\be}[1]{\begin{equation}\label{#1}}
 \newcommand{\ee}{\end{equation}}
 \newcommand{\bea}{\begin{eqnarray}}
 \newcommand{\eea}{\end{eqnarray}}
 \def\disp{\displaystyle}

 \def\gsim{ \lower .75ex \hbox{$\sim$} \llap{\raise .27ex \hbox{$>$}} }
 \def\lsim{ \lower .75ex \hbox{$\sim$} \llap{\raise .27ex \hbox{$<$}} }

 \begin{titlepage}

 \begin{flushright}
 arXiv:1010.1074
 \end{flushright}

 \title{\Large \bf Cosmological Constraints on
 the Sign-Changeable Interactions}

 \author{Hao~Wei\,}
 \email[\,email address:\ ]{haowei@bit.edu.cn}
 \affiliation{School of Physics, Beijing Institute
 of Technology, Beijing 100081, China}

 \begin{abstract}\vspace{1cm}
 \centerline{\bf ABSTRACT}\vspace{2mm}
 Recently, Cai and Su [Phys.\ Rev.\  D {\bf 81}, 103514 (2010)]
 found that the sign of interaction $Q$ in the dark sector
 changed in the approximate redshift range of
 $0.45\,\lsim\, z\,\lsim\, 0.9$, by using a model-independent
 method to deal with the observational data. In fact, this
 result raises a remarkable problem, since most of the familiar
 interactions cannot change their signs in the whole cosmic
 history. Motivated by the work of Cai and Su, we have
 proposed a new type of interaction in a previous work
 [H.~Wei, Nucl.\ Phys.\  B {\bf 845}, 381 (2011)]. The key
 ingredient is the deceleration parameter $q$ in
 the interaction $Q$, and hence the interaction $Q$ can change
 its sign when our universe changes from deceleration ($q>0$)
 to acceleration ($q<0$). In the present work, we consider the
 cosmological constraints on this new type of sign-changeable
 interactions, by using the latest observational data. We find
 that the cosmological constraints on the model parameters are
 fairly tight. In particular, the key parameter $\beta$ can be
 constrained to a narrow range.
 \end{abstract}


 \maketitle

 \end{titlepage}

 \renewcommand{\baselinestretch}{1.1}


\section{Introduction}\label{sec1}

In the dark energy cosmology~\cite{r1}, the well-known
 cosmological coincidence problem has an important position.
 This problem asks: why are we living in an epoch in which the
 densities of dark energy and matter are comparable? Since
 their densities scale differently with the expansion of our
 universe, there should be some fine-tunings. To alleviate this
 coincidence problem, it is natural to consider the possible
 interaction between dark energy and dark matter in the
 literature (see e.g.~\cite{r2,r3,r4,r5,r6,r7,r8,r9,r10,r11,r12}).
 In fact, since the natures of both dark energy and dark matter
 are still unknown, there is no physical argument to exclude
 the possible interaction between them. On the contrary, some
 observational evidences of the dark sector interaction have
 been found recently~\cite{r13,r14}. In the literature, it is
 usual to assume that dark energy and dark matter interact
 through a coupling term $Q$, according to
 \bea
 &&\dot{\rho}_m+3H\rho_m=Q\,,\label{eq1}\\
 &&\dot{\rho}_{de}+3H(\rho_{de}+p_{de})=-Q\,,\label{eq2}
 \eea
 where $\rho_m$ and $\rho_{de}$ are the densities of dark
 matter and dark energy (we assume that the baryon component
 can be ignored); $p_{de}$ is the pressure of dark energy;
 $H\equiv\dot{a}/a$ is the Hubble parameter; $a=(1+z)^{-1}$
 is the scale factor (we have set $a_0=1$; the subscript ``0''
 indicates the present value of corresponding quantity; $z$ is
 the redshift); a dot denotes the derivative with respect to
 cosmic time $t$. Note that Eqs.~(\ref{eq1}) and~(\ref{eq2})
 preserve the total energy conservation equation
 $\dot{\rho}_{tot}+3H(\rho_{tot}+p_{tot})=0$,
 where $\rho_{tot}=\rho_m+\rho_{de}$. Since there is {\em no}
 natural guidance from fundamental physics on the interaction
 $Q$, {\em one can only discuss it to a phenomenological
 level.} This is the realistic status of the interacting dark
 energy models so far.

The most familiar interactions extensively considered in the
 literature (see e.g.~\cite{r2,r3,r4,r5,r6,r7,r8,r9,r10,r11,r12})
 include $Q=3\alpha H\rho_m$, $Q=3\beta H\rho_{tot}$,
 and $Q=3\eta H\rho_{de}$. It is easy to see that these
 interactions are always positive or negative and hence cannot
 give the possibility to change their signs in the whole cosmic
 history. However, recently Cai and Su~\cite{r15} found that
 the sign of interaction $Q$ changed in the approximate
 redshift range of $0.45\,\lsim\, z\,\lsim\, 0.9$, by using a
 model-independent method to deal with the observational data.
 Obviously, this result raises a remarkable problem. Motivated
 by the work of Cai and Su, we have proposed a new type of
 interaction in a previous work~\cite{r16}, which is given by
 \be{eq3}
 Q=q(\alpha\dot{\rho}+3\beta H\rho)\,,
 \ee
 where $\alpha$ and $\beta$ are both dimensionless constants;
 the energy density $\rho$ could be $\rho_m$, $\rho_{tot}$ and
 $\rho_{de}$ for examples; the deceleration parameter
 \be{eq4}
 q\equiv -\frac{\ddot{a}}{aH^2}=-1-\frac{\dot{H}}{H^2}\,.
 \ee
 Obviously, this new type of interaction $Q$ can change its sign
 when our universe changes from deceleration ($q>0$) to
 acceleration ($q<0$). In fact, the deceleration parameter $q$
 in $Q$ is the key ingredient of this new interaction, which
 makes our proposal different from the previous ones considered
 in the literature. Note that the term $\alpha\dot{\rho}$ in
 $Q$ is introduced from the dimensional point of view (we refer
 to~\cite{r16} for details). One can remove this term by
 setting $\alpha=0$, and then $Q$ becomes simply
 $Q=3\beta qH\rho$ (in fact this is the very case which will
 be considered in the followings).

Since the appearance of the deceleration parameter $q$ in the
 interaction $Q$ looks speculative to some extent, we would
 like to say some words before going further. Firstly, as
 is well known, in the literature there is {\em no} natural
 guidance from fundamental physics on the interaction $Q$, one
 can only discuss it to a phenomenological level. In this sense,
 the other familiar interactions extensively considered in the
 literature have {\em no} better origin from the fundamental
 physics than the one proposed in Eq.~(\ref{eq3}). Secondly,
 we note that $q=-1-\dot{H}/H^2$ from Eq.~(\ref{eq4}) and
 $H^2 \propto\rho_{tot}$ from the Friedmann equation. Thus,
 one can regard the deceleration parameter
 $q=f(\rho_{tot},\dot{\rho}_{tot})$ as a function of the
 total energy density $\rho_{tot}=\rho_m+\rho_{de}$ and its
 derivative. In this sense, the interaction
 $Q=q(\alpha\dot{\rho}+3\beta H\rho)=f(\rho,\dot{\rho})$ is
 not so unusual, since it is reasonable to image that $Q$
 depends on the energy densities of dark energy and matter.
 Finally, while the familiar interactions extensively
 considered in the literature (such as $Q=3\alpha H\rho_m$,
 $Q=3\beta H\rho_{tot}$, and $Q=3\eta H\rho_{de}$) cannot
 give the possibility to change their signs in the whole
 cosmic history, our proposal in Eq.~(\ref{eq3}) provides a
 possible way out. So, we consider that it deserves further
 investigation.

In~\cite{r16}, we have studied the cosmological evolution
 of quintessence and phantom with this new type
 of sign-changeable interactions, and found some interesting
 results. In the present work, we would like to consider the
 cosmological constraints on this new type of sign-changeable
 interactions, by using the latest observational data. For
 simplicity, in this work, we restrict ourselves to the
 decaying $\Lambda$ model (see e.g.~\cite{r17} and references
 therein), namely, the role of dark energy is played by the
 decaying vacuum energy. In this case, Eq.~(\ref{eq2}) becomes
 \be{eq5}
 \dot{\rho}_\Lambda=-Q\,.
 \ee
 The Friedmann and Raychaudhuri equations are given by
 \bea
 &&H^2=\frac{\kappa^2}{3}\rho_{tot}=
 \frac{\kappa^2}{3}(\rho_\Lambda+\rho_m)\,,\label{eq6}\\
 &&\dot{H}=-\frac{\kappa^2}{2}(\rho_{tot}+p_{tot})=
 -\frac{\kappa^2}{2}\rho_m\,,\label{eq7}
 \eea
 where $\kappa^2\equiv 8\pi G$. Notice that we consider a flat
 Friedmann-Robertson-Walker (FRW) universe throughout this
 work. In Sec.~\ref{sec2}, we briefly introduce the latest
 observational data which will be used in this work. In
 Sec.~\ref{sec3}, we consider the cosmological constraints on
 three particular sign-changeable interactions, i.e.,
 \bea
 &&Q=q(\alpha\dot{\rho}_m+3\beta H\rho_m)\,,\label{eq8}\\
 &&Q=q(\alpha\dot{\rho}_{tot}+3\beta H\rho_{tot})\,,\label{eq9}\\
 &&Q=q(\alpha\dot{\rho}_\Lambda+3\beta H\rho_\Lambda)\,.\label{eq10}
 \eea
 Finally, some brief concluding remarks are given
 in Sec.~\ref{sec4}.


\section{Observational data}\label{sec2}

In the present work, we will consider the latest
 cosmological observations, namely, the 557 Union2 Type~Ia
 Supernovae (SNIa) dataset~\cite{r18}, the shift parameter
 $R$ from the Wilkinson Microwave Anisotropy Probe 7-year
 (WMAP7) data~\cite{r19}, and the distance parameter $A$ of
 the measurement of the baryon acoustic oscillation (BAO)
 peak in the distribution of SDSS luminous red
 galaxies~\cite{r20,r21}.

The data points of the 557 Union2 SNIa compiled
 in~\cite{r18} are given in terms of the distance modulus
 $\mu_{obs}(z_i)$. On the other hand, the theoretical
 distance modulus is defined as
 \be{eq11}
 \mu_{th}(z_i)\equiv 5\log_{10}D_L(z_i)+\mu_0\,,
 \ee
 where $\mu_0\equiv 42.38-5\log_{10}h$ and $h$ is the Hubble
 constant $H_0$ in units of $100~{\rm km/s/Mpc}$, whereas
 \be{eq12}
 D_L(z)=(1+z)\int_0^z \frac{d\tilde{z}}{E(\tilde{z};{\bf p})}\,,
 \ee
 in which $E\equiv H/H_0$, and ${\bf p}$ denotes the model
 parameters. Correspondingly, the $\chi^2$ from the 557
 Union2 SNIa is given by
 \be{eq13}
 \chi^2_{\mu}({\bf p})=\sum\limits_{i}\frac{\left[
 \mu_{obs}(z_i)-\mu_{th}(z_i)\right]^2}{\sigma^2(z_i)}\,,
 \ee
 where $\sigma$ is the corresponding $1\sigma$ error. The parameter
 $\mu_0$ is a nuisance parameter but it is independent of the data
 points. One can perform an uniform marginalization over $\mu_0$.
 However, there is an alternative way. Following~\cite{r22,r23}, the
 minimization with respect to $\mu_0$ can be made by expanding the
 $\chi^2_{\mu}$ of Eq.~(\ref{eq13}) with respect to $\mu_0$ as
 \be{eq14}
 \chi^2_{\mu}({\bf p})=\tilde{A}-2\mu_0\tilde{B}+\mu_0^2\tilde{C}\,,
 \ee
 where
 $$\tilde{A}({\bf p})=\sum\limits_{i}\frac{\left[\mu_{obs}(z_i)
 -\mu_{th}(z_i;\mu_0=0,{\bf p})\right]^2}
 {\sigma_{\mu_{obs}}^2(z_i)}\,,$$
 $$\tilde{B}({\bf p})=\sum\limits_{i}\frac{\mu_{obs}(z_i)
 -\mu_{th}(z_i;\mu_0=0,{\bf p})}{\sigma_{\mu_{obs}}^2(z_i)}\,,
 ~~~~~~~~~~
 \tilde{C}=\sum\limits_{i}\frac{1}{\sigma_{\mu_{obs}}^2(z_i)}\,.$$
 Eq.~(\ref{eq14}) has a minimum for
 $\mu_0=\tilde{B}/\tilde{C}$ at
 \be{eq15}
 \tilde{\chi}^2_{\mu}({\bf p})=
 \tilde{A}({\bf p})-\frac{\tilde{B}({\bf p})^2}{\tilde{C}}\,.
 \ee
 Since $\chi^2_{\mu,\,min}=\tilde{\chi}^2_{\mu,\,min}$
 obviously, we can instead minimize $\tilde{\chi}^2_{\mu}$
 which is independent of $\mu_0$.

There are some other relevant observational data, such as
 the observations of cosmic microwave background (CMB)
 anisotropy~\cite{r19} and large-scale structure
 (LSS)~\cite{r20}. However, using the full data of CMB and LSS
 to perform a global fitting consumes a large amount of
 computation time and power. As an alternative, one can
 instead use the shift parameter $R$ from the CMB, and the
 distance parameter $A$ of the measurement of the BAO peak in
 the distribution of SDSS luminous red galaxies. In the
 literature, the shift parameter $R$ and the distance parameter
 $A$ have been used extensively. It is argued that they are
 model-independent~\cite{r24}, while $R$ and $A$ contain the
 main information of the observations of CMB and BAO, respectively.

As is well known, the shift parameter $R$ of the CMB is defined
 by~\cite{r24,r25}
 \be{eq16}
 R\equiv\Omega_{m0}^{1/2}\int_0^{z_\ast}
 \frac{d\tilde{z}}{E(\tilde{z})}\,,
 \ee
 where $\Omega_{m0}$ is the present fractional energy density
 of pressureless matter; the redshift of recombination
 $z_\ast=1091.3$ which has been updated in the Wilkinson
 Microwave Anisotropy Probe 7-year (WMAP7) data~\cite{r19}. The
 shift parameter $R$ relates the angular diameter distance to
 the last scattering surface, the comoving size of the sound
 horizon at $z_\ast$ and the angular scale of the
 first acoustic peak in CMB power spectrum of temperature
 fluctuations~\cite{r24,r25}. The value of $R$ has been updated
 to $1.725\pm 0.018$ from the WMAP7 data~\cite{r19}. On the
 other hand, the distance parameter $A$ of the measurement of
 the BAO peak in the distribution of SDSS luminous red
 galaxies is given by~\cite{r20}
 \be{eq17}
 A\equiv\Omega_{m0}^{1/2}E(z_b)^{-1/3}\left[\frac{1}{z_b}
 \int_0^{z_b}\frac{d\tilde{z}}{E(\tilde{z})}\right]^{2/3},
 \ee
 where $z_b=0.35$. In~\cite{r21}, the value of $A$ has been
 determined to be $0.469\,(n_s/0.98)^{-0.35}\pm 0.017$. Here
 the scalar spectral index $n_s$ is taken to be $0.963$, which
 has been updated from the WMAP7 data~\cite{r19}. So, the total
 $\chi^2$ is given by
 \be{eq18}
 \chi^2=\tilde{\chi}^2_{\mu}+\chi^2_{CMB}+\chi^2_{BAO}\,,
 \ee
 where $\tilde{\chi}^2_{\mu}$ is given in Eq.~(\ref{eq15}),
 $\chi^2_{CMB}=(R-R_{obs})^2/\sigma_R^2$ and
 $\chi^2_{BAO}=(A-A_{obs})^2/\sigma_A^2$. The best-fit model
 parameters are determined by minimizing the total $\chi^2$.
 As in~\cite{r26,r27}, the $68.3\%$ confidence level is
 determined by
 $\Delta\chi^2\equiv\chi^2-\chi^2_{min}\leq 1.0$, $2.3$ and
 $3.53$ for $n_p=1$, $2$ and $3$, respectively, where $n_p$ is
 the number of free model parameters. Similarly, the $95.4\%$
 confidence level is determined by
 $\Delta\chi^2\equiv\chi^2-\chi^2_{min}\leq 4.0$, $6.17$ and
 $8.02$ for $n_p=1$, $2$ and $3$, respectively.


\section{Cosmological constraints on the sign-changeable
 interactions}\label{sec3}

In this section, we consider the cosmological constraints on
 the sign-changeable interactions given in
 Eqs.~(\ref{eq8})---(\ref{eq10}), by using the observational
 data mentioned in the previous section.


 \begin{center}
 \begin{figure}[tbhp]
 \centering
 \includegraphics[width=0.5\textwidth]{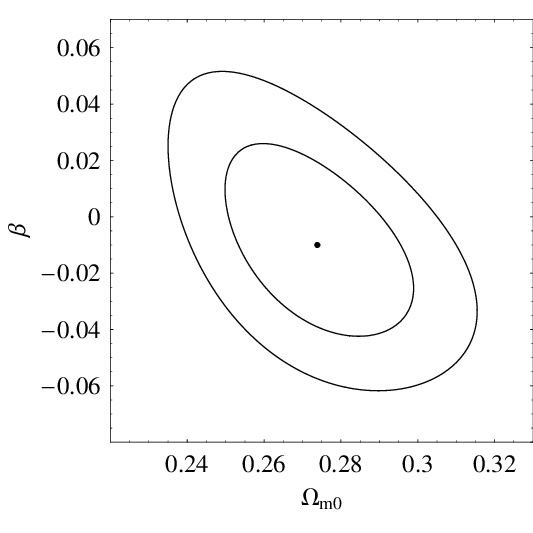}
 \caption{\label{fig1} The $68.3\%$ and $95.4\%$ confidence
 level contours in the $\Omega_{m0}-\beta$ plane for the case
 of $Q=3\beta qH\rho_m$. The best-fit parameters are also
 indicated by a black solid point.}
 \end{figure}
 \end{center}


\vspace{-11mm}  


\subsection{The case of
 $Q=q(\alpha\dot{\rho}_m+3\beta H\rho_m)$}\label{sec3a}

Firstly, we consider the case of
 $Q=q(\alpha\dot{\rho}_m+3\beta H\rho_m)$ given in Eq.~(\ref{eq8}).
 Substituting it into Eq.~(\ref{eq1}), one can find that
 \be{eq19}
 \dot{\rho}_m=\frac{\beta q-1}{1-\alpha q}\cdot 3H\rho_m\,.
 \ee
 Then, substituting into Eq.~(\ref{eq8}), we can finally obtain
 \be{eq20}
 Q=\frac{\beta-\alpha}{1-\alpha q}\cdot 3qH\rho_m\,.
 \ee
 From Eq.~(\ref{eq7}), we have
 \be{eq21}
 \rho_m=-\frac{2}{\kappa^2}\dot{H}\,.
 \ee
 Substituting into Eq.~(\ref{eq19}), we find that
 \be{eq22}
 \ddot{H}=\frac{\beta q-1}{1-\alpha q}\cdot 3H\dot{H}\,,
 \ee
 which is in fact a second-order differential equation for
 $H(t)$. We can change the time $t$ to scale factor $a$ with
 the help of the universal relation $\dot{f}=Haf^\prime$ for
 any function $f$ (where a prime denotes the derivative with
 respect to scale factor $a$), and recast Eq.~(\ref{eq22}) as
 \be{eq23}
 aH^{\prime\prime}+\frac{a}{H}H^{\prime\,2}+H^\prime=
 \frac{\beta q-1}{1-\alpha q}\cdot 3H^\prime\,,
 \ee
 which is a second-order differential equation for $H(a)$.
 Note that the deceleration parameter
 \be{eq24}
 q=-1-\frac{\dot{H}}{H^2}=-1-\frac{a}{H}H^\prime\,,
 \ee
 is also a function of $H$ and $H^\prime$. Unfortunately,
 if $\alpha\not=0$, there is {\em no} analytical solution for
 the second-order  

\newpage  


 \begin{center}
 \begin{figure}[tbhp]
 \centering
 \includegraphics[width=1.0\textwidth]{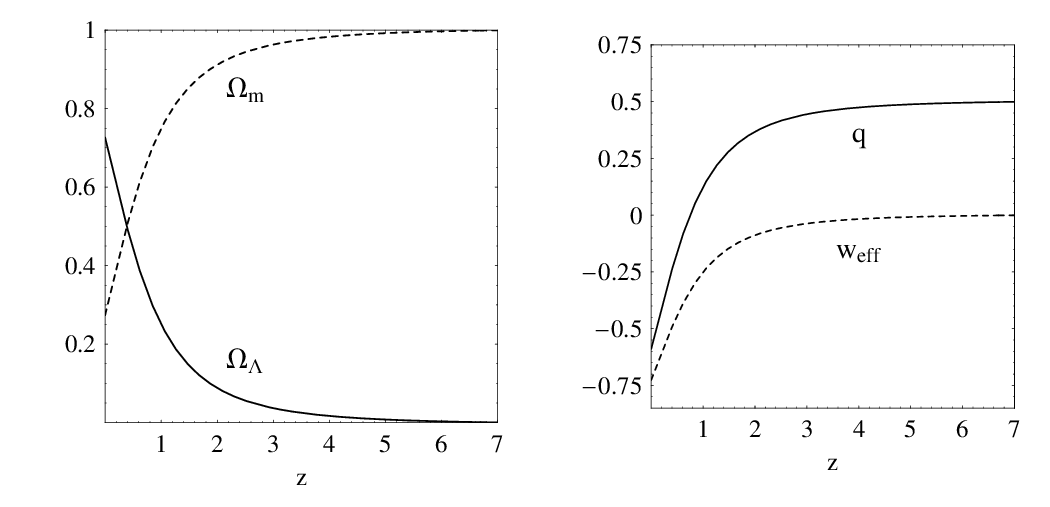}
 \caption{\label{fig2} $\Omega_m$, $\Omega_\Lambda$, $q$ and
 $w_{\rm eff}$ as functions of redshift $z$ with the best-fit
 parameters for the case of $Q=3\beta qH\rho_m$.}
 \end{figure}
 \end{center}


\vspace{-5mm}  

\noindent differential equation~(\ref{eq23}), because
 one will encounter a transcendental equation. Therefore, we
 consider only the case of $\alpha=0$ in this work. In this
 case, the sign-changeable interaction reads
 \be{eq25}
 Q=3\beta qH\rho_m\,.
 \ee
 By solving the second-order differential equation~(\ref{eq23})
 with $\alpha=0$, we find that
 \be{eq26}
 H(a)=C_{12}\left[\,3C_{11}(1+\beta)-(2+3\beta)
 \,a^{-3(1+\beta)}\,\right]^{1/(2+3\beta)},
 \ee
 where $C_{11}$ and $C_{12}$ are both integral constants, which
 can be determined in the following. From Eq.~(\ref{eq21}), we
 find that the fractional energy density of dark matter is given by
 \be{eq27}
 \Omega_m\equiv \frac{\kappa^2 \rho_m}{3H^2}
 =-\frac{2\dot{H}}{3H^2}=-\frac{2aH^\prime}{3H}\,.
 \ee
 Substituting Eq.~(\ref{eq26}) into Eq.~(\ref{eq27}), we have
 \be{eq28}
 \Omega_m=\frac{2\left(1+\beta\right)}{2+3\beta
 -3C_{11}\left(1+\beta\right)\,a^{3\left(1+\beta\right)}}\,.
 \ee
 Requiring $\Omega_m(a=1)=\Omega_{m0}$, we obtain
 \be{eq29}
 C_{11}=\frac{\Omega_{m0}(2+3\beta)
 -2(1+\beta)}{3\Omega_{m0}(1+\beta)}\,.
 \ee
 On the other hand, requiring $H(a=1)=H_0$, from
 Eq.~(\ref{eq26}) we can find that
 \be{eq30}
 C_{12}=H_0\left[\,3C_{11}(1+\beta)
 -(2+3\beta)\,\right]^{-1/(2+3\beta)}.
 \ee
 Substituting Eqs.~(\ref{eq29}) and (\ref{eq30})
 into Eq.~(\ref{eq26}), we finally obtain
 \be{eq31}
 E\equiv\frac{H}{H_0}=\left\{1-
 \frac{2+3\beta}{2(1+\beta)}\,\Omega_{m0}\left[1-
 (1+z)^{3(1+\beta)}\right]\right\}^{1/(2+3\beta)}.
 \ee
 There are two free model parameters, namely $\Omega_{m0}$ and
 $\beta$. Note that when $\beta=0$, Eq.~(\ref{eq31}) reduces
 to $E(z)=\left[\Omega_{m0}(1+z)^3+\left(1-\Omega_{m0}\right)
 \right]^{1/2}$, i.e., the one of $\Lambda$CDM model.

By minimizing the corresponding total $\chi^2$
 in Eq.~(\ref{eq18}), we find the best-fit parameters
 $\Omega_{m0}=0.2738$ and $\beta=-0.010$, whereas
 $\chi^2_{min}=542.725$. In Fig.~\ref{fig1}, we present
 the corresponding $68.3\%$ and $95.4\%$ confidence level
 contours in the $\Omega_{m0}-\beta$ plane. Obviously, the
 current observational data slightly prefer a negative~$\beta$.
 We are also interested to the fractional energy densities
 $\Omega_m$ given in Eq.~(\ref{eq28}) and
 $\Omega_\Lambda=1-\Omega_m$, the deceleration parameter $q$
 given in Eq.~(\ref{eq24}), and the effective equation-of-state
 parameter (EoS) $w_{\rm eff}\equiv p_{tot}/\rho_{tot}=(2q-1)/3$.
 We present them as functions of redshift $z$ with the best-fit
 model parameters in Fig.~\ref{fig2}. It is easy to find the
 transition redshift $z_t=0.7489$ where the universe changes
 from deceleration ($q>0$) to acceleration ($q<0$). Since the
 best-fit $\beta$ is negative, dark matter decays into dark
 energy ($Q<0$) when $z>z_t$, and dark energy decays into dark
 matter ($Q>0$) when $z<z_t$. The interaction $Q$ crosses the
 non-interacting line ($Q=0$) at $z_t$.


 \begin{center}
 \begin{figure}[tbhp]
 \centering
 \includegraphics[width=0.5\textwidth]{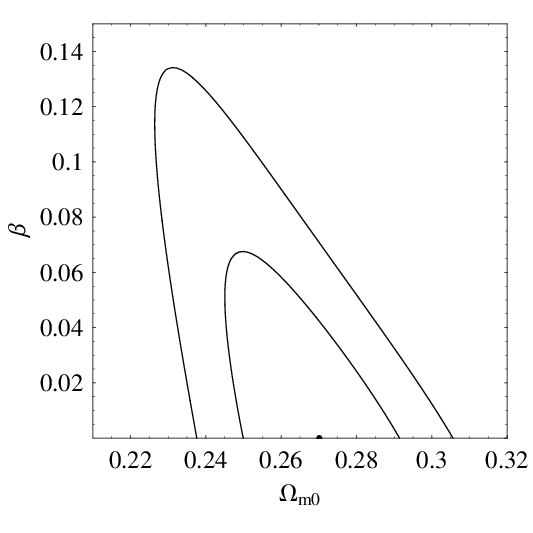}
 \caption{\label{fig3} The same as in Fig.~\ref{fig1}, but
 for the case of $Q=3\beta qH\rho_{tot}$ with the condition
 $\beta\geq 0$.}
 \end{figure}
 \end{center}


\vspace{-11mm}  


 \begin{center}
 \begin{figure}[tbhp]
 \centering
 \includegraphics[width=1.0\textwidth]{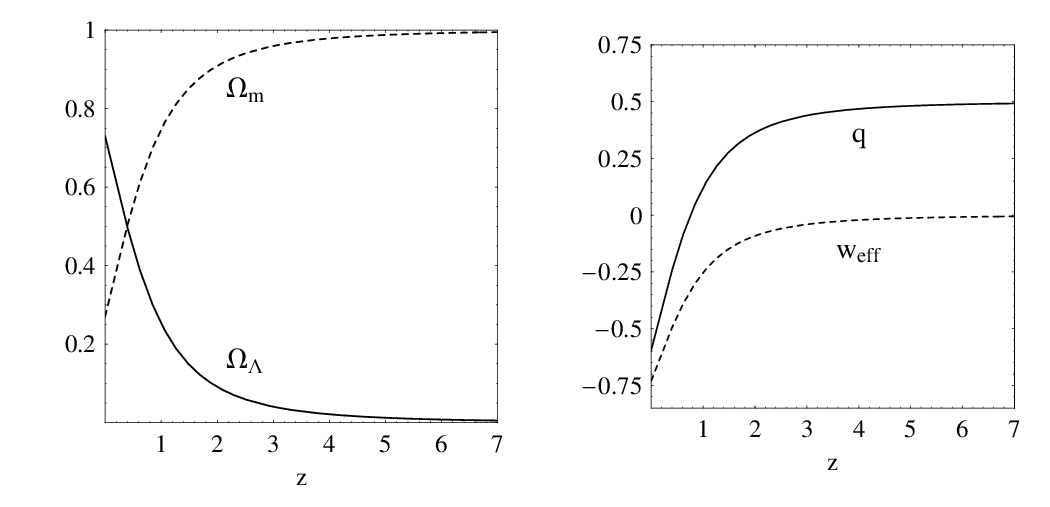}
 \caption{\label{fig4} The same as in Fig.~\ref{fig2}, but
 for the case of $Q=3\beta qH\rho_{tot}$ with the condition
 $\beta\geq 0$.}
 \end{figure}
 \end{center}


\vspace{-11mm}  


\subsection{The case of
 $Q=q(\alpha\dot{\rho}_{tot}+3\beta H\rho_{tot})$}\label{sec3b}

Secondly, we consider the case of
 $Q=q(\alpha\dot{\rho}_{tot}+3\beta H\rho_{tot})$ given in
 Eq.~(\ref{eq9}). From Eq.~(\ref{eq6}), it is easy to find
 $\rho_{tot}=3H^2/\kappa^2$. Substituting into Eq.~(\ref{eq9}),
 we can finally obtain
 \be{eq32}
 Q=\frac{3qH^3}{\kappa^2}\left(2\alpha\frac{\dot{H}}{H^2}
 +3\beta\right).
 \ee
 Substituting Eqs.~(\ref{eq21}) and (\ref{eq32})
 into Eq.~(\ref{eq1}), we have
 \be{eq33}
 \ddot{H}+3H\dot{H}\left(1+\alpha q\right)+
 \frac{9}{2}\beta qH^3=0\,.
 \ee
 Similarly, we recast it as
 \be{eq34}
 aH^{\prime\prime}+\frac{a}{H}H^{\prime\,2}+
 \left(4+3\alpha q\right)H^\prime+\frac{9\beta qH}{2a}=0\,,
 \ee
 which is a second-order differential equation for $H(a)$.
 Note that the deceleration parameter $q$ is also a function
 of $H$ and $H^\prime$ [cf. Eq.~(\ref{eq24})]. Similar to the
 case of $Q=q(\alpha\dot{\rho}_m+3\beta H\rho_m)$, we consider
 only the case of $\alpha=0$ in this work. In this case, the
 sign-changeable interaction reads
 \be{eq35}
 Q=3\beta qH\rho_{tot}\,.
 \ee
 By solving the second-order differential equation~(\ref{eq34})
 with $\alpha=0$, we find that
 \be{eq36}
 H(a)=C_{22}\cdot a^{-3(2-3\beta+r_1)/8}
 \cdot\left(a^{3r_1/2}+C_{21}\right)^{1/2},
 \ee
 where $C_{21}$, $C_{22}$ are both integral constants, and
 \be{eq37}
 r_1\equiv\sqrt{4+\beta\left(4+9\beta\right)}\,.
 \ee
 Substituting Eq.~(\ref{eq36}) into Eq.~(\ref{eq27}), we have
 \be{eq38}
 \Omega_m=\frac{1}{4}\left[\,2-3\beta
 +\left(\frac{2C_{21}}{a^{3r_1/2}+C_{21}}-1\right)r_1\right].
 \ee
 Requiring $\Omega_m(a=1)=\Omega_{m0}$, we obtain
 \be{eq39}
 C_{21}=-1+\frac{2\,r_1}{2-3\beta-4\Omega_{m0}+r_1}\,.
 \ee
 On the other hand, requiring $H(a=1)=H_0$, from
 Eq.~(\ref{eq36}) we can find that
 \be{eq40}
 C_{22}=H_0\left(1+C_{21}\right)^{-1/2}.
 \ee
 From Eqs.~(\ref{eq36}) and (\ref{eq40}), it is easy to obtain
 \be{eq41}
 E\equiv\frac{H}{H_0}=(1+z)^{3(2-3\beta+r_1)/8}\cdot
 \left[\frac{(1+z)^{-3r_1/2}+C_{21}}{1+C_{21}}\right]^{1/2},
 \ee
 where $C_{21}$ and $r_1$ have been given in Eqs.~(\ref{eq39})
 and (\ref{eq37}), respectively. There are two free model
 parameters, namely $\Omega_{m0}$ and $\beta$. Note that when
 $\beta=0$, Eq.~(\ref{eq41}) reduces to $E(z)=\left[\Omega_{m0}
 (1+z)^3+\left(1-\Omega_{m0}\right)\right]^{1/2}$, i.e., the
 one of $\Lambda$CDM model.

Imposing the condition $0\leq\Omega_m\leq 1$ when $a\to 0$, we have
 $\beta\geq 0$ from Eq.~(\ref{eq38}). Under this condition, by
 minimizing the corresponding total $\chi^2$ in Eq.~(\ref{eq18}),
 we find the best-fit parameters $\Omega_{m0}=0.2701$ and
 $\beta=0.0$, whereas $\chi^2_{min}=542.919$. In
 Fig.~\ref{fig3}, we present the corresponding $68.3\%$ and
 $95.4\%$ confidence level contours in the $\Omega_{m0}-\beta$
 plane. In Fig.~\ref{fig4}, we also present the $\Omega_m$
 given in Eq.~(\ref{eq38}), $\Omega_\Lambda=1-\Omega_m$, $q$
 given in Eq.~(\ref{eq24}) and
 $w_{\rm eff}\equiv p_{tot}/\rho_{tot}=(2q-1)/3$ as functions
 of redshift $z$ with the best-fit model parameters. The
 universe changes from deceleration ($q>0$) to acceleration
 ($q<0$) at the transition redshift $z_t=0.7549$.

\begin{center}
 \begin{figure}[tbp]
 \centering
 \includegraphics[width=0.5\textwidth]{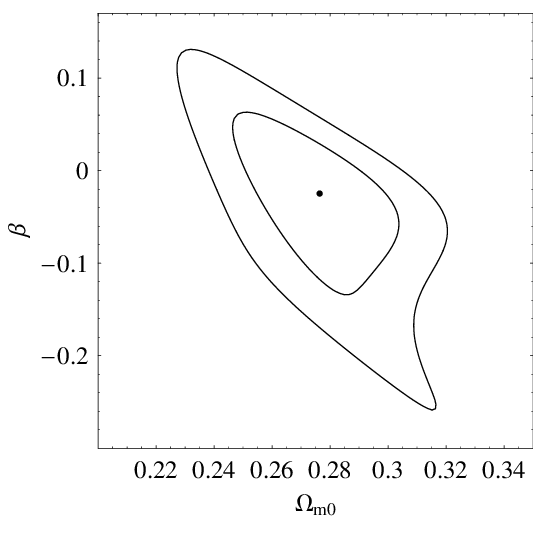}
 \caption{\label{fig5} The same as in Fig.~\ref{fig1}, but
 for the case of $Q=3\beta qH\rho_{tot}$ without the condition
 $\beta\geq 0$.}
 \end{figure}
 \end{center}

\begin{center}
 \begin{figure}[tbp]
 \centering
 \includegraphics[width=1.0\textwidth]{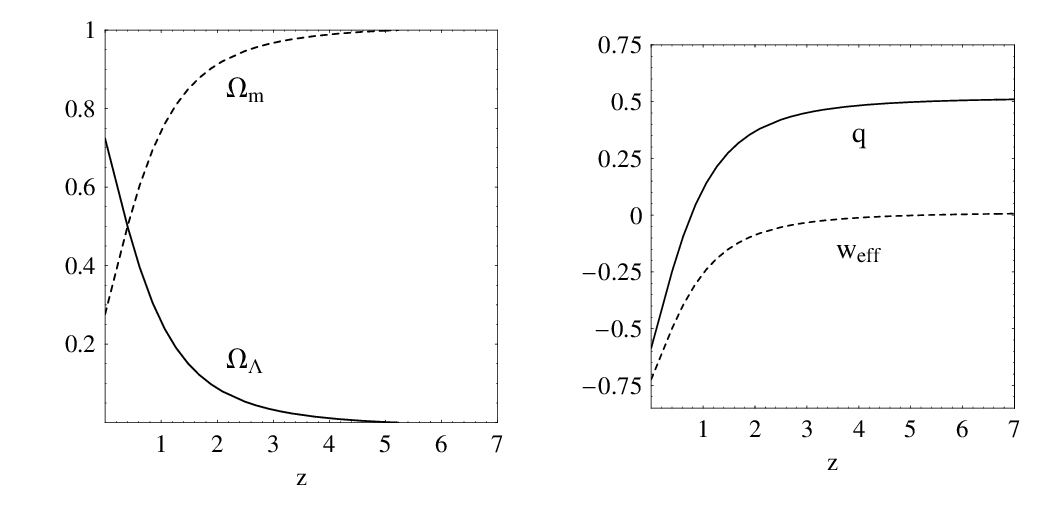}
 \caption{\label{fig6} The same as in Fig.~\ref{fig2}, but
 for the case of $Q=3\beta qH\rho_{tot}$ without the condition
 $\beta\geq 0$.}
 \end{figure}
 \end{center}

However, the above best-fit model with $\beta=0$ is in fact
 the $\Lambda$CDM model without interaction between dark energy
 and dark matter. So, we would like to give up the condition
 $\beta\geq 0$. This means that in the early universe we
 have $\Omega_m\geq 1$ and then $\Omega_\Lambda\leq 0$, namely,
 $\rho_\Lambda$ might be negative. Since the negative energy
 density can arise in quantum field theory (see e.g.~\cite{r28}
 for a good review), it is reasonable to consider this possibility.
 Without the condition $\beta\geq 0$, by minimizing the
 corresponding total $\chi^2$ in Eq.~(\ref{eq18}), we find the
 best-fit parameters $\Omega_{m0}=0.2764$ and $\beta=-0.0247$,
 whereas $\chi^2_{min}=542.711$. In Fig.~\ref{fig5}, we present
 the corresponding $68.3\%$ and $95.4\%$ confidence level
 contours in the $\Omega_{m0}-\beta$ plane. Obviously, the
 current observational data slightly prefer a negative $\beta$.
 In Fig.~\ref{fig6}, we also present the $\Omega_m$ given in
 Eq.~(\ref{eq38}), $\Omega_\Lambda=1-\Omega_m$, $q$ given in
 Eq.~(\ref{eq24}) and
 $w_{\rm eff}\equiv p_{tot}/\rho_{tot}=(2q-1)/3$ as functions
 of redshift~$z$ with the best-fit model parameters. It is
 easy to find the transition redshift $z_t=0.7688$ where the
 universe changes from deceleration ($q>0$) to acceleration
 ($q<0$). Since the best-fit $\beta$ is negative, dark matter
 decays into dark energy ($Q<0$) when $z>z_t$, and dark energy
 decays into dark matter ($Q>0$) when $z<z_t$. The interaction
 $Q$ crosses the non-interacting line ($Q=0$) at $z_t$.


\vspace{-11mm}  


\vspace{-6mm}  


\subsection{The case of
 $Q=q(\alpha\dot{\rho}_\Lambda+3\beta H\rho_\Lambda)$}\label{sec3c}

Finally, we consider the case of
 $Q=q(\alpha\dot{\rho}_\Lambda+3\beta H\rho_\Lambda)$ given
 in Eq.~(\ref{eq10}). Substituting it into Eq.~(\ref{eq5}),
 one can find that
 \be{eq42}
 \dot{\rho}_\Lambda=-\frac{3\beta qH\rho_\Lambda}{1+\alpha q}\,.
 \ee
 Then, substituting into Eq.~(\ref{eq10}), we can finally
 obtain
 \be{eq43}
 Q=\frac{3\beta qH\rho_\Lambda}{1+\alpha q}\,.
 \ee
 From Eqs.~(\ref{eq6}) and (\ref{eq7})
 [or equivalently Eq.~(\ref{eq21})], we have
 \be{eq44}
 \rho_\Lambda=\frac{3}{\kappa^2}H^2-\rho_m=
 \frac{1}{\kappa^2}\left(3H^2+2\dot{H}\right).
 \ee
 Substituting Eqs.~(\ref{eq21}), (\ref{eq43}) and (\ref{eq44})
 into Eq.~(\ref{eq1}), we find that
 \be{eq45}
 \ddot{H}+3H\dot{H}\left(1+\frac{\beta q}{1+\alpha q}\right)
 +\frac{9\beta qH^3}{2(1+\alpha q)}=0\,.
 \ee
 Similarly, we recast it as
 \be{eq46}
 aH^{\prime\prime}+\frac{a}{H}H^{\prime\,2}
 +\left(4+\frac{3\beta q}{1+\alpha q}\right)H^\prime+
 \frac{9\beta qH}{2a(1+\alpha q)}=0\,,
 \ee
 which is a second-order differential equation for $H(a)$.
 Note that the deceleration parameter $q$ is also a function
 of $H$ and $H^\prime$ [cf. Eq.~(\ref{eq24})]. Unfortunately,
 if $\alpha\not=0$, there is {\em no} analytical solution for
 the second-order differential equation~(\ref{eq46}), because
 one will encounter a transcendental equation. Therefore, we
 consider only the case of $\alpha=0$ in this work. In this
 case, the sign-changeable interaction reads
 \be{eq47}
 Q=3\beta qH\rho_\Lambda\,.
 \ee
 By solving the second-order differential equation~(\ref{eq46})
 with $\alpha=0$, we find that
 \be{eq48}
 H(a)=C_{32}\cdot a^{-3(2-5\beta+r_2)/[4(2-3\beta)]}\cdot
 \left(a^{3r_2/2}+C_{31}\right)^{1/(2-3\beta)},
 \ee
 where $C_{31}$, $C_{32}$ are both integral constants, and
 \be{eq49}
 r_2\equiv\sqrt{\left(2-\beta\right)^2}=\left|\,2-\beta\,\right|\,.
 \ee
 Substituting Eq.~(\ref{eq48}) into Eq.~(\ref{eq27}), we have
 \be{eq50}
 \Omega_m=\frac{1}{2\left(2-3\beta\right)}\left[2-5\beta+
 \left(\frac{2C_{31}}{a^{3r_2/2}+C_{31}}-1\right)r_2\right]\,.
 \ee
 Requiring $\Omega_m(a=1)=\Omega_{m0}$, we obtain
 \be{eq51}
 C_{31}=-1+\frac{2\,r_2}{2-5\beta+
 r_2+2\Omega_{m0}\left(3\beta-2\right)}\,.
 \ee
 On the other hand, requiring $H(a=1)=H_0$, from
 Eq.~(\ref{eq48}) we get
 \be{eq52}
 C_{32}=H_0\left(1+C_{31}\right)^{1/(3\beta-2)}.
 \ee
 From Eqs.~(\ref{eq48}) and (\ref{eq52}), it is easy to obtain
 \be{eq53}
 E\equiv\frac{H}{H_0}=(1+z)^{3\left(2-5\beta+r_2\right)/\left[
 4\left(2-3\beta\right)\right]}\cdot\left[\frac{(1+z)^{-3r_2/2}
 +C_{31}}{1+C_{31}}\right]^{1/\left(2-3\beta\right)},
 \ee
 where $C_{31}$ and $r_2$ have been given in Eqs.~(\ref{eq51})
 and (\ref{eq49}), respectively. There are two free model
 parameters, namely $\Omega_{m0}$ and $\beta$. Note that when
 $\beta=0$, Eq.~(\ref{eq53}) reduces to $E(z)=\left[\Omega_{m0}
 (1+z)^3+\left(1-\Omega_{m0}\right)\right]^{1/2}$, i.e., the
 one of $\Lambda$CDM model.


 \begin{center}
 \begin{figure}[tbp]
 \centering
 \includegraphics[width=0.5\textwidth]{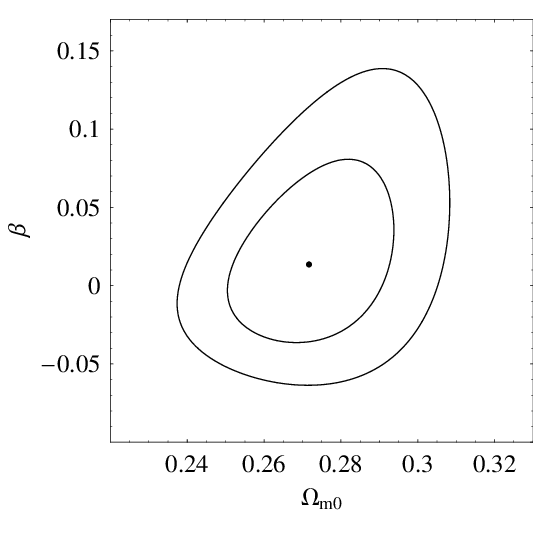}
 \caption{\label{fig7} The same as in Fig.~\ref{fig1}, but
 for the case of $Q=3\beta qH\rho_\Lambda$.}
 \end{figure}
 \end{center}


\vspace{-11mm}  

\begin{center}
 \begin{figure}[tbp]
 \centering
 \includegraphics[width=1.0\textwidth]{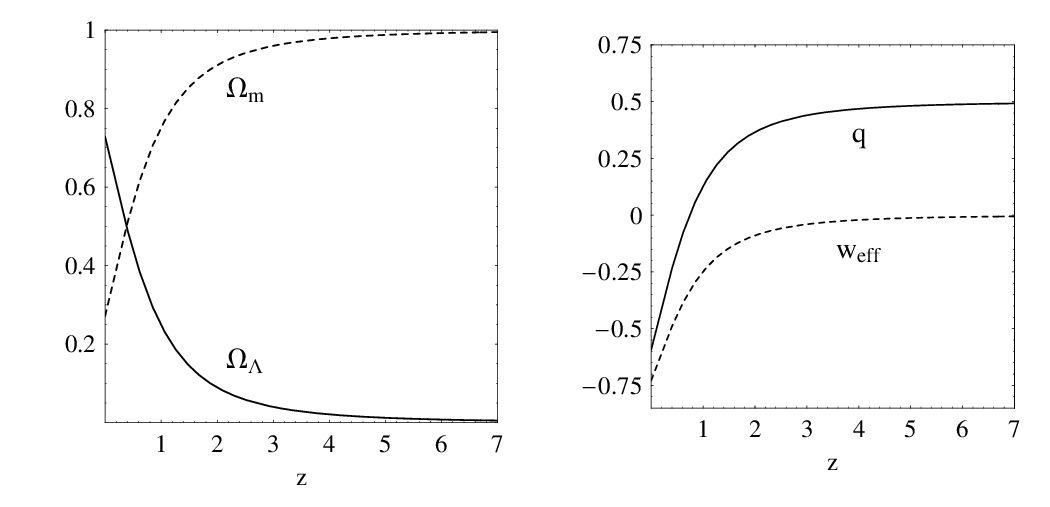}
 \caption{\label{fig8} The same as in Fig.~\ref{fig2}, but
 for the case of $Q=3\beta qH\rho_\Lambda$.}
 \end{figure}
 \end{center}

By minimizing the corresponding total $\chi^2$
 in Eq.~(\ref{eq18}), we find the best-fit parameters
 $\Omega_{m0}=0.2717$ and $\beta=0.0136$, whereas
 $\chi^2_{min}=542.778$. In Fig.~\ref{fig7}, we present
 the corresponding $68.3\%$ and $95.4\%$ confidence level
 contours in the $\Omega_{m0}-\beta$ plane. Obviously, the
 current observational data slightly prefer a positive $\beta$.
 In Fig.~\ref{fig8}, we also present the $\Omega_m$ given in
 Eq.~(\ref{eq50}), $\Omega_\Lambda=1-\Omega_m$, $q$ given in
 Eq.~(\ref{eq24}) and
 $w_{\rm eff}\equiv p_{tot}/\rho_{tot}=(2q-1)/3$ as functions
 of redshift $z$ with the best-fit model parameters. It is
 easy to find the transition redshift $z_t=0.7398$ where the
 universe changes from deceleration ($q>0$) to acceleration
 ($q<0$). Since the best-fit $\beta$ is positive, dark energy
 decays into dark matter ($Q>0$) when $z>z_t$, dark matter
 decays into dark energy ($Q<0$) when $z<z_t$. The interaction
 $Q$ crosses the non-interacting line ($Q=0$) at $z_t$.


\vspace{-11mm}  


\section{Concluding remarks}\label{sec4}

Recently, Cai and Su~\cite{r15} found that the sign of
 interaction $Q$ in the dark sector changed in the approximate
 redshift range of $0.45\,\lsim\, z\,\lsim\, 0.9$, by using a
 model-independent method to deal with the observational data.
 In fact, this result raises a remarkable problem, since most
 of the familiar interactions cannot change their signs in the
 whole cosmic history. Motivated by the work of Cai and Su, we
 have proposed a new type of interaction in a previous
 work~\cite{r16}. The key ingredient is the deceleration
 parameter $q$ in the interaction $Q$, and hence
 the interaction $Q$ can change its sign when our universe
 changes from deceleration ($q>0$) to acceleration ($q<0$). In
 the present work, we consider the cosmological constraints
 on this new type of sign-changeable interactions, by using the
 latest observational data. We find that the cosmological
 constraints on the model parameters are fairly tight. In
 particular, the key parameter $\beta$ has been constrained to
 a narrow range.

Some remarks are in order. Firstly, we briefly consider the
 comparison of these models. For convenience, we also consider
 the well-known $\Lambda$CDM model in addition. In fact, it
 corresponds to the decaying $\Lambda$ model with $Q=0$.
 Fitting $\Lambda$CDM model to the observational data
 considered in the present work, it is easy to find the
 corresponding best-fit parameter $\Omega_{m0}=0.2701$, whereas
 $\chi^2_{min}=542.919$. Of course, we would like to also
 consider the decaying $\Lambda$ model with a traditional
 interaction $Q=3\beta H\rho_m$ which cannot change its sign
 in the whole cosmic history. The corresponding $E(z)$ can be
 found in e.g.~\cite{r12}, namely
 \be{eq54}
 E(z)=\left[\frac{\Omega_{m0}}{1-\beta}(1+z)^{3(1-\beta)}+
 \left(1-\frac{\Omega_{m0}}{1-\beta}\right)\right]^{1/2}.
 \ee
 Fitting to the same observational data, we find the best-fit
 parameters $\Omega_{m0}=0.2731$ and $\beta=-0.0021$, whereas
 $\chi^2_{min}=542.735$. A conventional criterion for model
 comparison in the literature is $\chi^2_{min}/dof$, in which
 the degree of freedom $dof=N-k$, whereas $N$ and $k$ are the
 number of data points and the number of free model parameters,
 respectively. We present the $\chi^2_{min}/dof$ for all the 6
 models in Table~\ref{tab1}. On the other hand, there are other
 criterions for model comparison in the literature, such as
 Bayesian Information Criterion (BIC) and Akaike Information
 Criterion (AIC). The BIC is defined by~\cite{r29,r31}
 \be{eq55}
 {\rm BIC}=-2\ln{\cal L}_{max}+k\ln N\,,
 \ee
 where ${\cal L}_{max}$ is the maximum likelihood. In the
 Gaussian cases, $\chi^2_{min}=-2\ln{\cal L}_{max}$. So, the
 difference in BIC between two models is given by
 $\Delta{\rm BIC}=\Delta\chi^2_{min}+\Delta k \ln N$. The AIC
 is defined by~\cite{r30,r31}
 \be{eq56}
 {\rm AIC}=-2\ln{\cal L}_{max}+2k\,.
 \ee
 The difference in AIC between two models is given by
 $\Delta{\rm AIC}=\Delta\chi^2_{min}+2\Delta k$.
 In Table~\ref{tab1}, we also present the $\Delta$BIC and
 $\Delta$AIC of all the 6 models considered in this work.
 Notice that $\Lambda$CDM has been chosen to be the fiducial
 model when we calculate $\Delta$BIC and $\Delta$AIC. From
 Table~\ref{tab1}, it is easy to see that the rank of models
 is coincident in all the 3 criterions ($\chi^2_{min}/dof$,
 BIC and AIC). The $\Lambda$CDM model is still the best one.
 However, it is well known that $\Lambda$CDM model is plagued
 with the cosmological constant problem and the coincidence
 problem (see e.g.~\cite{r1}). On the other hand, there are
 some observational evidences for the interaction between dark
 energy and dark matter~\cite{r13,r14}, and the coincidence
 problem can be alleviated in the interacting dark energy
 models. Therefore, it is still worthwhile to study the
 interacting dark energy models. Although the model with
 traditional interaction (which cannot change its sign) is very
 close to the other models with sign-changeable interactions,
 the latter are phenomenally richer (see e.g.~\cite{r16}).
 Therefore, we consider that the models with sign-changeable
 interactions deserve further investigations.


 \begin{table}[tbp]
 \begin{center}
 \begin{tabular}{llllllll}\hline\hline
 Model & $\Lambda$CDM & $Q=3\beta H\rho_m$~~~~
 & $Q=3\beta qH\rho_m$~~~ & $Q=3\beta qH\rho_{tot}$~~~
 & $Q=3\beta qH\rho_{tot}$~~~ & $Q=3\beta qH\rho_\Lambda$ \\
 & ($Q=0$) & & & with $\beta\geq 0$ & without $\beta\geq 0$
 & \\[0.5mm] \hline
 Best fit & $\Omega_{m0}=0.2701$~~ & $\Omega_{m0}=0.2731$
 & $\Omega_{m0}=0.2738$ & $\Omega_{m0}=0.2701$
 & $\Omega_{m0}=0.2764$ & $\Omega_{m0}=0.2717$ \\
 & & $\beta=-0.0021$ & $\beta=-0.010$ & $\beta=0.0$
 & $\beta=-0.0247$ & $\beta=0.0136$\\ \hline
 $\chi^2_{min}$ & 542.919 & 542.735 & 542.725 & 542.919
 & 542.711 & 542.778 \\
 $k$ & 1 & 2 & 2 & 2 & 2 & 2\\
 $\chi^2_{min}/dof~~$  & 0.9730 & 0.9744 & 0.9744 & 0.9747
 & 0.9743 & 0.9745 \\
 $\Delta$BIC & 0 & 6.142 & 6.132 & 6.326 & 6.118 & 6.185 \\
 $\Delta$AIC & 0 & 1.816 & 1.806 & 2.0 & 1.792 & 1.859 \\
 Rank & 1 & 4 & 3 & 6 & 2 & 5 \\
 \hline\hline
 \end{tabular}
 \end{center}
 \caption{\label{tab1} Summarizing all the 6 models considered
 in this work.}
 \end{table}


Secondly, we note that the case of $Q=3\beta qH\rho_\Lambda$
 is fairly different from the cases of $Q=3\beta qH\rho_m$ and
 $Q=3\beta qH\rho_{tot}$. Comparing Fig.~\ref{fig7} with
 Figs.~\ref{fig1}, \ref{fig3} and \ref{fig5}, it is easy to
 see that the direction of contours in the $\Omega_{m0}-\beta$
 plane is rightward for the case of $Q=3\beta qH\rho_\Lambda$,
 whereas the ones are leftward for the cases of
 $Q=3\beta qH\rho_m$ and $Q=3\beta qH\rho_{tot}$.
 From Table~\ref{tab1}, we find that the best-fit $\beta$ is
 positive for the case of $Q=3\beta qH\rho_\Lambda$, whereas
 the ones are negative (or zero) for the cases of
 $Q=3\beta qH\rho_m$ and $Q=3\beta qH\rho_{tot}$. This means
 that in the case of $Q=3\beta qH\rho_\Lambda$ the interaction
 $Q$ crosses the non-interacting line ($Q=0$) from above to
 below, whereas in the cases of $Q=3\beta qH\rho_m$ and
 $Q=3\beta qH\rho_{tot}$ the interaction $Q$ crosses the
 non-interacting line ($Q=0$) from below to above. This is
 physically interesting, because $Q>0$ means that the energy
 transfers from dark energy to dark matter, whereas $Q<0$ means
 that the energy transfers from dark matter to dark energy.

Finally, in this work the role of dark energy is only played
 by the decaying $\Lambda$ (vacuum energy), whereas the
 parameter $\alpha$ in the sign-changeable interactions are
 chosen to be zero. So, the constraints obtained in this work
 cannot be directly used to the models different from the
 ones considered here. In fact, the interacting dark energy
 models with sign-changeable interactions can be generalized.
 For instance, one can choose dark energy to be the one
 with a constant or variable EoS (including parameterized EoS,
 or even the ones of quintessence, phantom, k-essence,
 Chaplygin gas, quintom, hessence, holographic or agegraphic
 dark energy, and so on). Of course, one can also let the
 parameter $\alpha$ be free and then constrain the models
 numerically.


\section*{ACKNOWLEDGEMENTS}
We thank the referee for quite useful comments and suggestions,
 which help us to improve this work. We are grateful to
 Professors Rong-Gen~Cai and Shuang~Nan~Zhang for helpful
 discussions. We also thank Minzi~Feng, as well
 as Xiao-Peng~Ma, for kind help and discussions. This work was
 supported in part by NSFC under Grant No.~10905005, the
 Excellent Young Scholars Research Fund of Beijing Institute
 of Technology, and the Fundamental Research Fund of Beijing
 Institute of Technology.

\renewcommand{\baselinestretch}{1.2}


\end{document}